\documentclass[preprint,12pt,3p]{elsarticle}

\usepackage{amssymb}
\usepackage{graphicx}
\usepackage[cmex10]{amsmath}
\usepackage{array}
\usepackage{mdwmath}
\usepackage{caption}

\journal{Optical Fiber Technology}

\begin{document}

\begin{frontmatter}
	
	\title{What comes after optical-bypass network? A study on optical-computing-enabled network}

	\author[label1]{Dao Thanh Hai \corref{cor1}}
	\address[label1]{School of Science, Engineering and Technology, RMIT University Vietnam}
	
	\cortext[cor1]{Corresponding author}
	
	\ead{hai.dao5@rmit.edu.vn}
	
	
	
	\begin{abstract}
		A new architectural paradigm, named, optical-computing-enabled network, is proposed as a potential evolution of the currently used optical-bypass framework. The main idea is to leverage the optical computing capabilities performed on transitional lightpaths at intermediate nodes and such proposal reverses the conventional wisdom in optical-bypass network, that is, separating in-transit lightpaths in avoidance of unwanted interference. In optical-computing-enabled network, the optical nodes are therefore upgraded from conventional functions of add-drop and cross-connect to include optical computing / processing capabilities. This is enabled by exploiting the superposition of in-transit lightpaths for computing purposes to achieve greater capacity efficiency. While traditional network design and planning algorithms have been well-developed for optical-bypass framework in which the routing and resource allocation is dedicated to each optical channel (lightpath), more complicated problems arise in optical-computing-enabled architecture as a consequence of intricate interaction between optical channels and hence resulting into the establishment of the so-called integrated / computed lightpaths. This necessitates for a different framework of network design and planning to maximize the impact of optical computing opportunities. In highlighting this critical point, a detailed case study exploiting the optical aggregation operation to re-design the optical core network is investigated in this paper. Optical aggregation enables the combination of lower-speed and/or lower-order format channels into a single higher-rate and/or higher-order format one for saving spectrum resources and such new perspective give rises to major challenges involving aggregation assignments. Specifically, the determination of demands for aggregation, the nodes at which the optical aggregation takes place and the wavelength selection for aggregated lightpaths constitute a critical issue to be tackled. Numerical results obtained from extensive simulations on the COST239 network are presented to quantify the efficacy of optical-computing-enabled approach versus the conventional optical-bypass-enabled one. 
	\end{abstract}
	
	
\end{frontmatter}


\section{Introduction}
\label{intro}

Internet traffic has been remaining a continued explosive growth with an estimated rate of $30\%$ annually. It is therefore projected that in the next ten years, a roughly 14-fold increase in the network capacity will be expected \cite{futureoptics1, futureoptics2}. In addressing such a relentless traffic increase, on one hand, the long-term solution is clearly to expand the system capacity \cite{crunch, crunch2, intro1, intro2}. On the other hand, as network operators will not likely spend 14 times more, achieving economies of scales by lowering the cost per bit both in capital and operational expenditures will be an equally important goal. In this context, several innovative proposals from both technological and architectural front have been studied, investigated and eventually brought into play with wide deployments \cite{intro3, trend1, trend2, trend3}. With the goal of scaling up optical fiber capacities, advances in optical transmissions have resulted in multi-band, multi-dimensional and spatial-division-multiplexing technologies and these advances have been culminated in a recently demonstrated world-record by National Institute of Information and Communications Technology (NICT) Japan reaching 319 Tb/s over 3000 km \cite{NICT}. It is noted that such an achievement has represented more than a 300-fold increases in a span of a quarter century from the 1 Tb/s record made in 1996 \cite{20years}. While technological approaches are centered on extending the system capacity by orders of magnitude, an important aspect coupled with the rise of system capacity is to harness the available transmission capacity through optical networking and bandwidth management \cite{hai_iet, hai_icact, hai_csndsp, hai_wiley, hai_nics, hai_atc}. This is where architectural solutions come into play with the aim to reduce the effective traffic load and to manage the bandwidth directly in the optical layer \cite{all-optical, hai_oft3, hai_oft2, hai_optik}.  \\

From an architectural perspective, optical networking has had a paradigm shift in the time around the year 2000s with the proposal and then quickly coming to wide adoption of optical-bypass networking \cite{all-optical}. Enabling technologies for optical-bypass networking including progressively more complex and more automated re-configurable optical add-drop multiplexing (ROADM) nodes, long-haul transmission and advanced networking algorithms have been built, developed and matured in the last two decades to optimally harness the system capacity \cite{nodearchitecture, Algorithm4, Algorithm5}. In an optical-bypass network, in-transit lightpaths are optically cross-connected from one end to the other end, and therefore the major benefits associated with optical-bypass operations compared to its predecessor (i.e., optical-electrical-optical mode) is the elimination of costly and power-hungry high-speed electronic interfaces for an optical channel en route from a source to destination. However, the fundamental assumption behind the optical-bypass framework lies in the fact that when a lightpath crosses an intermediate node, it must be separated from other transitional lightpaths at that node, usually in time, frequency or spatial dimensions in avoidance of unwanted interference which is considered to deteriorate signal qualities \cite{all-optical, nodearchitecture}. This turns out to be a bottleneck in pushing forward the use of photonic technologies for greater scalability and flexibility in handling explosive Internet traffic growth. \\

Inspired by the recently massive investments and consequently rapid progress in photonic computing technologies permitting the purpose-controlled interference between optical channels for various computing capabilities \cite{nature, nature2, nature3, photonicmit1, photonicmit2, optical_processing_5, xor3}, this paper proposes a new architectural paradigm for next-generation optical networks, entitled, optical-computing-enabled network, as a potential evolution of the currently used optical-bypass framework. The main idea is to leverage the optical mixing operations between transitional lightpaths at intermediate nodes and such proposal reverses the conventional wisdom in optical-bypass network, that is, separating in-transit lightpaths in avoidance of unwanted interference. In optical-computing-enabled network, the optical nodes are therefore upgraded from ordinary functions of add-drop and cross-connect to include optical computing capabilities performed on lightpath entities. Note that in our previous works \cite{hai_tnsm, hai_springer5, hai_mttw22, hai_mttw21, hai_springer3, hai_wrap, hai_systems, hai_oft, hai_comcom, hai_comcom2, hai_comletter, hai_access, hai_rtuwo, hai_springer, hai_springer2}, we coined the term, optical-processing-enabled network, encompassing new scenarios where optical nodes are enhanced with various all-optical signal processing capabilities that are performed on both individual lightpaths (e.g., wavelength / modulation format conversion) and/or group of favorable lightpath entities (e.g., optical XOR between lightpath entities). Although optical-processing-enabled and optical-computing-enabled network could be used interchangeably, the latter one is particularly inclined to the vision where many computing functions are pushed down to the optical layer, leveraging the photon's bandwidth and energy-efficiency, paving the way for the era of optical-layer intelligence. With optical-computing-enabled network, we envision the seamless integration of optical computing and communications infrastructure, potentially driving the promising developments of in-network optical computing that can sustain the bandwidth demand for the future connectivity spanning from connected people, things to connected intelligence. \\

While traditional network design and planning algorithms have been well-developed for optical-bypass framework where the routing and resource allocation is dedicated to each optical channel \cite{Algorithm5, hai_springer4, hai_iet, hai_thesis, hai_ps1, hai_ps2, hai_sigtel1, hai_icist2}, more complicated problems arise in optical-computing-enabled architecture as a consequence of intricate interaction between optical channels and hence the establishment of the so-called integrated / computed lightpaths \cite{aina_arxiv, hai_springer5, hai_ro}. Indeed, new network design problems are emerging from the application of photonic computing operations between transitional lightpaths and this necessitates a different framework of network design and planning algorithms in order to maximize the impact of such lightpath superposition. In highlighting this critical point, a detailed case study exploiting the optical aggregation operation to re-design the optical transport networks is investigated in this paper. Optical aggregation enables the combination of lower-speed and/or lower-order format channels into a single higher-rate and/or higher-order format for saving spectrum resources \cite{apl, agg1, agg2, agg5, agg13} and such new perspective give rises to major challenges involving aggregation assignments. Specifically, the determination of demands for aggregation, the node at which the aggregation takes place and the wavelength selection for the aggregated lightpaths constitute a critical issue to be tackled. Numerical results simulated on the COST239 network are presented to quantify the efficacy of optical-computing-enabled approach versus the conventional optical-bypass-enabled one. It is worthy to mention that related works in the literature concerning the efficient utilization of optical aggregation for network-wide benefits have been gaining momentum and leading industrial players, notably INFINERA, have been promoting this promising technique in the context of point-to-multipoint (P2MP) optical networks thanks to digital sub-carrier multiplexing-based (DSCM) technologies \cite{p2mp0, p2mp1}. While DSCM-based technologies may be ripe for the cost-efficient realization of optical aggregation, they may not be the efficient platform for implementation of other advanced optical computing operations. Our proposed vision is indeed broader in scope, foreseeing various optical computing operations at the lightpath scale that could be exploited and realized for next-generation optical networks and how such added capabilities are translated to network-wide gains. We then focus on the optical aggregation as a case study for illustrative purpose of our proposed vision. \\

The remainder of the paper is organized as follows. The subsequent part, Section 2, introduces the concept of an optical-computing-enabled paradigm and how crucially different it is in comparison with its counterpart, an optical-bypass one. In this section, as a case study, we present the functional operation of the optical aggregation and highlight how such an operation could be exploited for more efficient traffic provisioning. In optimizing the network impact of optical aggregation, a new and more complicated network design problem entitled, routing, wavelength and aggregation assignment is introduced and mathematically formulated in the form of an integer linear programming model in Section 3. Numerical evaluations drawing an extensive comparison between our proposal that leverages the optical aggregation within the framework of optical-computing-enabled networks and the traditional optical-bypass networking is presented in Section 4. The comparison is obtained from the realistic COST239 network and all-to-one traffic setting. Finally, Section 5 summarizes the paper and touch on potential future works.

\section{Optical-computing-enabled Network with Optical Aggregation}

Photonic signal processing / computing technologies have been accelerating in recent years thanks to massive investments, heralding the renaissance of optical computing \cite{nature, nature2, nature3, opticalprocessing0, opticalprocessing1, opticalprocessing2}. Basically, various signal processing functions could be realized directly in optical domain thanks to exploiting nonlinear optical processes, for example, four-wave mixing, self-phase modulation, cross-phase modulation, occurring in optical materials such as highly nonlinear fibers and silicon waveguides when propagating one or many optical channels simultaneously \cite{optical_processing_1, optical_processing_5}. In addition to traditional signal processing operations like logical computing, linear transformation and convolution multiplication \cite{opticalprocessing0} that have gone through many breakthroughs recently thanks to AI accelerator developments, new scenarios encompassing unseen signal processing and computing capabilities are envisioned by processing photons by photons directly in the optical domain. Specifically, exploiting the superposition of optical channels at intermediate nodes for various signal processing / computing purposes remains vastly unexplored as this represents a paradigm shift compared to the widely used optical-bypass network where each optical channel is kept optically intact from a source to a destination. This section focuses on the effective usage of optical aggregation function whose enabling technologies have been progressing fast lately, in an attempt to reduce the effective traffic load in the network and consequently boost the system capacity. \\ 

Merging lower-speed optical channels into a single higher-speed one has been the major function in designing and operating optical networks \cite{apl, agg1, agg2}. From the functional perspective, an optical aggregator supports adding two or more lower bit-rate and/or lower-order modulation format optical channels into a single higher bit-rate and/or higher-order modulation format one. In the following illustrative example, we examine the use of an optical aggregator whose the main function is to add two QPSK signals to a single 16-QAM channel \cite{apl, optical_processing_2} as depicted by the schematic diagram in Fig. 1. The goal of doing so is to pack more data into a given spectrum channel and thereby improving the spectral efficiency. \\

\begin{figure}[!ht]
	\centering
	\includegraphics[width=0.7\linewidth]{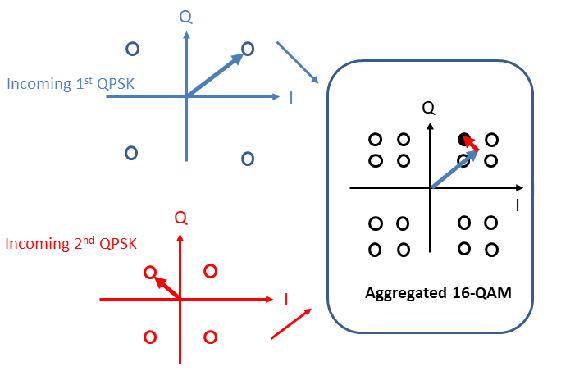}
	\caption{Schematic Diagram for Optical Aggregation}
	\label{fig:topology}
\end{figure}

\begin{figure}[!ht]
	\centering
	\includegraphics[width=\linewidth, height = 7cm]{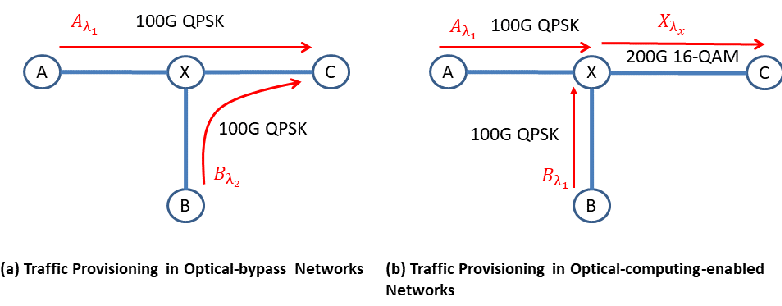}
	\caption{Traffic Provisioning in Optical-bypass vs. Optical-computing-enabled Networks}
	\label{fig:provisioning}
\end{figure}

Assuming that there are two traffic demands of the same $100G$ from node $A$ and node $B$ to node $C$ respectively, as shown in Fig. 2. We first consider the conventional way of accommodating these traffic demands in optical-bypass networking and one way of doing so is shown in Fig. 2(a) where each demand is served with a lightpath (i.e., $A_{\lambda_1}$ and $B_{\lambda_2}$) composed of a route and a wavelength. It is important to note that in optical-bypass networking, when the lightpath  $A_{\lambda_1}$ and $B_{\lambda_2}$ are routed over the intermediate node $X$, they have to be separated in the frequency domain in order to avoid interference and on the common output fiber link $XC$, due to the wavelength uniqueness constraint, two wavelengths are therefore needed. We then turn the attention to the case where the optical aggregation is enabled at node $X$. By enabling the optical aggregation, the two 100G QPSK transitional lightpaths crossing node $X$ could optically be added together to generate the output signal of 200G which is modulated on 16-QAM format. It can be observed from the Fig. 2(b) that by having a single wavelength channel of 200G capacity, greater capacity efficiency has been realized. At the common destination node $C$, the aggregated lightpath (a.k.a., the integrated / computed lightpath) could be decomposed into constituent ones and such a decomposition operation could be performed either in the optical or electrical domain. From a network design perspective, the fundamental difference between an optical-computing-enabled network and optical-bypass one is the emergence of integrated / computed lightpaths (i.e., $X_{\lambda_x}$) as the consequence of interaction between transitional lightpaths for computing purposes. Managing and optimizing such newly appeared lightpaths thus constitute a critical issue to be addressed in maximizing the aggregation opportunities and consequently unlocking the potential benefits of the new optical-computing-enabled mode. In this context, new network design and planning algorithms should be developed to optimally determine the pairing of demands for aggregation, the respective aggregation node and the transmission parameters for aggregated lightpaths. In the Section that follows, we formulate and develop an optimal algorithm for designing and planning optical networks that tap into optical aggregation capabilities. \\

\section{A Mathematical Formulation for the Routing, Wavelength and Aggregation Assignment Problem}
Different from the traditional routing and wavelength assignment in optical-bypass networking, more complicated problems arise in optical-computing-enabled networks with optical aggregation. In this section, we develop a mathematical model for solving the routing, wavelength and aggregation assignment problem arisen in the upgrading optical nodes functions with optical aggregation capabilities. For ease of operation, optical aggregation is permitted between two lightpaths of the same wavelength and the de-aggregation operation only takes place at the destination node. The formulation aims at the optimal design of optical-aggregation-enabled networks to support a given set of traffic demands with minimum number of used wavelengths. \\

\noindent{Inputs:}
\begin{footnotesize}
	\begin{itemize}
		\item $G(V,E)$: A directed graph consisting of $|V|$ nodes and $|E|$ links represents a fiber-optic network; $s(e)$ and $r(e)$ represents the beginning and ending node for a link $e \in E$, respectively.
		\item $D$: A set of requested traffic demands and each demand $d \in D$ is represented by its source node $s(d)$ and its destination node $r(d)$, respectively. Each demands $d \in D$ request the same \textit{one unit traffic} (e.g., 100G).
		\item $W$: A set of available wavelengths on each fiber link and the link capacity measured in number of wavelengths is $|W|$.  
	\end{itemize}
\end{footnotesize}

\noindent{Variables:}
\begin{footnotesize}
	\begin{itemize}
		\item $x_{e, w}^{d} \in \{0,1\} $: equals 1 if a link $e$ and a wavelength $w$ is used by a demand $d$, 0 otherwise.

		\item $\theta_{w}^{d} \in \{0,1\} $: equals 1 if a demand $d$ uses a wavelength $w$,  0 otherwise.
		
		\item $y_{e, w}^{d, v} \in \{0,1\} $: equals 1 if a demand $d$ using a wavelength $w$ is optically aggregated with another demand at a node $v$ and a link $e$ is selected for routing the resultant aggregated lightpath, 0 otherwise.
		
		\item $\delta_{v}^{d} \in \{0, 1\} $: equals 1 if $v$ is the aggregation node for a demand $d$, 0 otherwise.
		
		\item $f_{d_1}^{d_2} \in \{0, 1\} $: equals 1 if two demands $d_1$ and $d_2$ are aggregated with each other, 0 otherwise.
		
		\item $\gamma_{e,w} \in \{0, 1\}$: equals 1 if a wavelength $w$ is used on a link $e$, 0 otherwise. 
		\item $\alpha_{w} \in \{0, 1\}$: equals 1 if a wavelength $w$ is used (i.e., a wavelength is considered being used if it is occupied by a demand in any link of the network), 0 otherwise. \\
	\end{itemize}
\end{footnotesize}
\noindent{Objective function:}
\begin{footnotesize}
	\begin{equation} \label{eq:obj}
		\textit{Minimize} \; \sum_{w \in W}  \alpha_{w} \\
	\end{equation}
\end{footnotesize}

\noindent{Subject to the following conditions:}
\begin{footnotesize}
	\begin{equation}\label{eq:c1}
		\sum_{w \in W} {\theta^d_w} = 1 \; \; \forall d \in D 
	\end{equation}
	
	\begin{equation} \label{eq:c2}
		\begin{split}
			\sum_{e \in {E}: v\equiv s(e)} {x_{e, w}^{d}} -\sum_{e \in {E}: v \equiv r(e)} {x_{e, w}^{d}}= \\		
			\begin{cases} 
				\theta_{w}^{d} &\mbox{if } v \equiv s(d) \\ 
				-\theta_{w}^{d}& \mbox{if } v \equiv r(d)\\
				$0$ & otherwise \\
			\end{cases}     \qquad \qquad \forall v \in V, \forall d \in D, \forall w \in W \hfill
		\end{split}
	\end{equation}

	
	\begin{align} \label{eq:c4} 
		\begin{split}
			\sum_{d \in D} x_{e, w}^{d} - \frac{1}{2} \sum_{d \in D} \sum_{v \in V} {y_{e, w}^{d, v}} = \gamma_{e,w} \\
			\qquad \forall e \in E, \forall w \in W
		\end{split}
	\end{align}
	
	
	\begin{align} \label{eq:c6} {
			\sum_{v \in V} \delta_{v}^{d} \leq 1 \qquad and \qquad \delta_{v}^{d} = 0 \qquad \mbox{if } v \equiv r(d) \qquad \forall  d \in D
		}
	\end{align}
	
	\begin{align} \label{eq:c7} {
			\sum_{d_2 \in D} f_{d_1}^{d_2} \leq 1 \qquad \forall d_1 \in D
		}
	\end{align}
	
	\begin{equation} \label{eq:c8}
		{f^{d_1}_{d_1}} + \sum_{d_2 \in D: r(d_2) \neq r(d_1)}  {f^{d_1}_{d_2}} = 0 \qquad \forall d_1 \in D
	\end{equation}

	\begin{align} \label{eq:c9} {
			f_{d_1}^{d_2} = f_{d_2}^{d_1} \qquad \forall d_1, d_2 \in D
		}
	\end{align}

	\begin{align} \label{eq:c10} {
			\sum_{d_2 \in D} f_{d_1}^{d_2} = \sum_{v \in V} \delta_{v}^{d_1} \qquad \forall d_1 \in D
		}
	\end{align}
	
	\begin{align} \label{eq:c11} {
			\sum_{w \in W} \sum_{v \in V} y_{e, w}^{d_1, v} \leq \sum_{d_2 \in D} f_{d_1}^{d_2} \qquad \forall d_1 \in D, \forall e \in E
		}
	\end{align}
	
	\begin{align} \label{eq:c12} {
			\sum_{w \in W} y_{e, w}^{d, v}  \leq \delta_{v}^{d}   \qquad \forall e \in E, \forall v \in V, \forall d \in D
		}
	\end{align}

	%
	%
	\begin{align} \label{eq:c15} {
			\theta_{w_1}^{d_1} - \theta_{w_2}^{d_2}+f_{d_1}^{d_2} \leq 1 \qquad \forall d_1, d_2 \in D, \forall w_1, w_2 \in W
		}
	\end{align}
	
	\begin{align} \label{eq:c16} {
			\theta_{w_2}^{d_2} - \theta_{w_1}^{d_1}+f_{d_1}^{d_2} \leq 1 \qquad \forall d_1, d_2 \in D, \forall w_2, w_1 \in W
		}
	\end{align}
	
	\begin{align} \label{eq:c17} {
			\delta_{v}^{d_1} - \delta_{v}^{d_2}+f_{d_1}^{d_2} \leq 1 \qquad \forall d_1, d_2 \in D, \forall v \in V
		}
	\end{align}
	
	\begin{align} \label{eq:c18} {
			\delta_{v}^{d_2} - \delta_{v}^{d_1}+f_{d_1}^{d_2} \leq 1 \qquad \forall d_1, d_2 \in D, \forall v \in V
		}
	\end{align}
	
	\begin{align} \label{eq:c19} {
			y_{e, w}^{d, v}  \leq  x_{e, w}^{d}  \qquad \forall w \in W, \forall d \in D, \forall v \in V, \forall e \in E
		}
	\end{align}
	
	\begin{equation} \label{eq:c20}
		\begin{split}
			\sum_{w \in {W}} (\sum_{e \in E: i=s(e)} y_{e, w}^{d, v} - \sum_{e \in E: i=r(e)} y_{e, w}^{d, v})=\\
			\begin{cases} 
				\delta_{v}^{d} &\mbox{if } i \equiv v \\ 
				-\delta_{v}^{d} & \mbox{if } i \equiv r(d)\\
				0 & \mbox{otherwise}
			\end{cases} \qquad \qquad \forall d \in D, \forall v \in V, \forall i \in V \hfill
		\end{split}
	\end{equation}
	
	\begin{equation} \label{eq:c21}
		\sum_{e\in E} \gamma_{e,w} \leq |E| \times \alpha_{w}  \qquad \forall w \in W
	\end{equation}	
	
\end{footnotesize}

The objective function defined by Eq. \ref{eq:obj} is to minimize the number of used wavelengths. Constraints in Eq. \ref{eq:c1} guarantee the serving of all demands by finding the appropriate wavelength. The flow conservation is formulated in  Eq. \ref{eq:c2}. The constraint of wavelength uniqueness on each fiber link is ensured by Eq. \ref{eq:c4}. Constraints expressed by Eq. \ref{eq:c6} say that each traffic demand could be aggregated at one and only one intermediate node. The condition that each demand could be aggregated with at most one another demand of the same destination is indicated by constraints in Eq. \ref{eq:c7}, Eq. \ref{eq:c8} and Eq. \ref{eq:c9}. A set of conditions formulated by Eq. \ref{eq:c10}, Eq. \ref{eq:c11} and Eq. \ref{eq:c12} are for coherence among variables, i.e., if two traffic demands are aggregated, the respective aggregation node must be found and that aggregated lightpath must also find its respective route and wavelength. The same wavelength constraint for two demands that could be aggregated is captured by Eq. \ref{eq:c15} and Eq. \ref{eq:c16}. Constraints in Eq. \ref{eq:c17} and Eq. \ref{eq:c18} mean that if two demands are aggregated, such aggregation must take place at the same node. The coherent between the route for aggregated lightpath and the route of each individual demand is expressed by Eq. \ref{eq:c19}. The constraint in Eq. \ref{eq:c20} is the flow conservation of aggregated traffic. The final Eq. \ref{eq:c21} is the definition of using a wavelength on the network. \\

It is worth noticing that in addition to typical variables and constraints accounting for the selection of route and assigning wavelength for each demand, new variables and constraints account for the interaction of demands have been introduced. Specifically, the coming into play of new variables $y_{e, w}^{d, v}$ and constraints in Eq. \ref{eq:c15}, Eq. \ref{eq:c16}, Eq. \ref{eq:c19} cause the model one order of magnitude computationally harder than its counterpart, that is, the traditional routing and wavelength assignment in optical-bypass networking \cite{Algorithm4}. 

\section{Numerical Results}
\begin{figure}[!ht]
	\centering
	\includegraphics[width=0.55\linewidth, height=8cm]{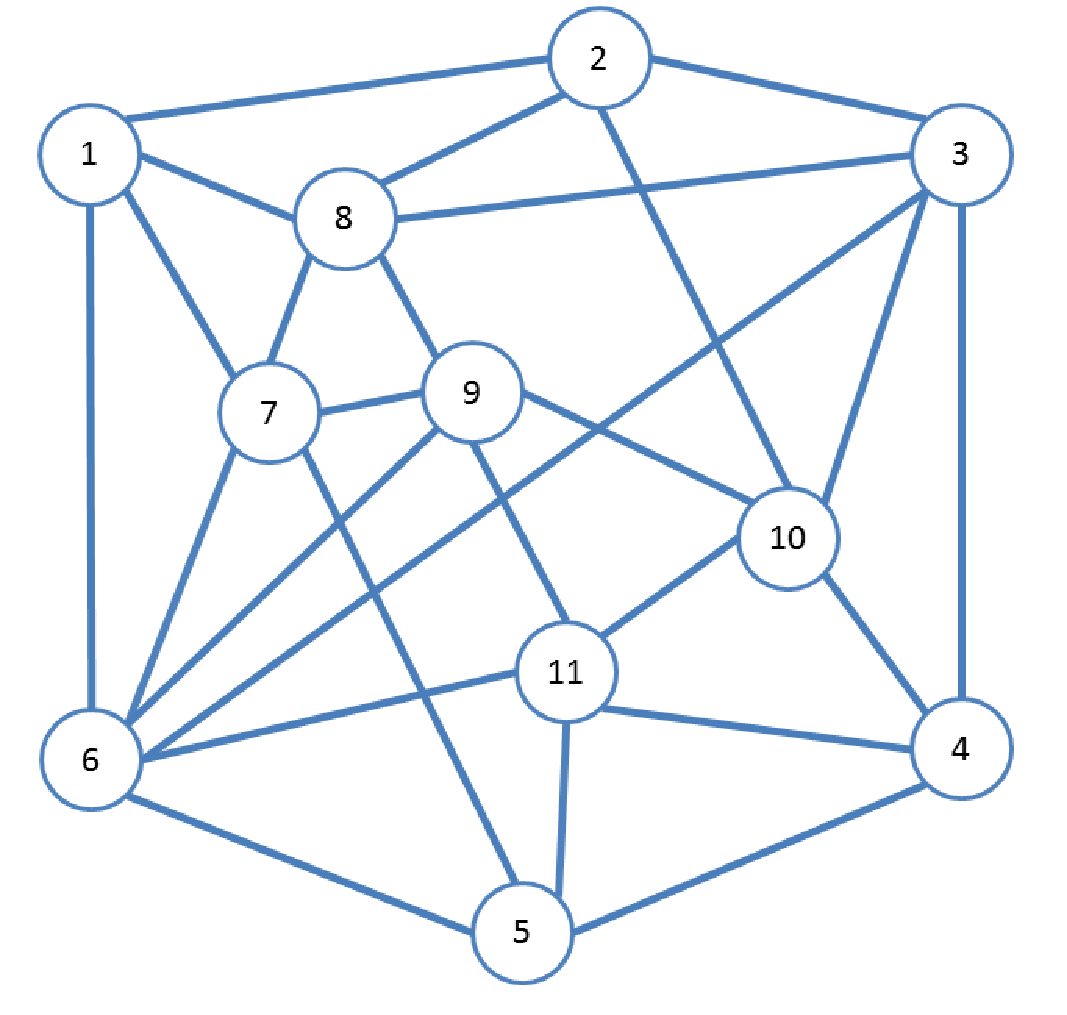}
	\caption{COST239 network}
	\label{fig:cost239}
\end{figure}

This section is dedicated to provide simulation results drawing on the performance comparison between the conventional optical-bypass design versus the optical-computing-enabled one with optical aggregation. The metric for comparison is the minimum number of used wavelengths to accommodate a given set of traffic demands. The simulation is performed on the realistic topology, COST239, composing of 11 nodes and 52 links and its nodes degree range from 4 to 6 as shown in Fig. \ref{fig:cost239}. In paving the favorable conditions for aggregation, the traffic under consideration is assumed to be all-to-one where one destination node is designated and there is a traffic request from all remaining nodes in the network to that designated destination node. The mathematical formulation underpinning the network designs was implemented in Matlab and then solved by CPLEX with the academic license. To guarantee a reliable and fair comparison, all the results for both designs are optimally collected.  \\

Table \ref{tab:result1} reports the optimal number of wavelengths needed for both two approaches at different nodal degrees. For both designs, it shows that the lower the destination's nodal degree is, the more wavelengths are needed to support the traffic demands. This could be explained by the fact that the lower the destination's nodal degree is, the less incoming fiber is connected to the node and therefore, due to the wavelength uniqueness constraint on each fiber, more wavelengths are required. Moreover, it is revealed that the introduction of optical aggregation indeed help to achieve a better performance than the traditional way for accommodating traffic demands in optical-bypass and such improvement is consistently realized across all destination nodes. This is to verify the efficacy of exploiting optical aggregation that could result in practical network-wide gain. It is worthy to note that the spectral saving enabled by the optical-computing-based approach comes at the cost of solving a more complicated network design problem to determine the optimal aggregation assignment when compared to its counterpart in optical-bypass framework. \\ 

\begin{table}[!ht]
	\caption{Performance Comparison}
	\label{tab:result1}
	\centering
	\begin{tabular}{ccc}
		\hline
		Node & Design & Number of Wavelengths\\
		\hline
		Degree 4 & Optical-bypass-based & 3 \\
		& Optical-computing-based & 2 \\
		\cline{1-3}
		Degree 5 & Optical-bypass-based & 2\\
		& Optical-computing-based & 1 \\
		\cline{1-3}
		Degree 6 & Optical-bypass-based & 2 \\
		& Optical-computing-based & 1 \\
		\cline{1-3}  	       			
		\hline
	\end{tabular}
\end{table}

\begin{table}[!ht]
	\caption{Routing and Wavelength Assignment Information in Optical-bypass-enabled Architecture}
	\label{tab:result2}
	\centering
	\begin{tabular}{ccc}
		\hline
		Traffic Demands & Routing & $\lambda$ \\
		\hline
		(2$\rightarrow$1) & 2-1 &   1 \\
		(3$\rightarrow$1) & 3-8-1 & 2 \\
		(4$\rightarrow$1) & 4-10-2-1 &  2 \\
		(5$\rightarrow$1) & 5-7-1 &  3 \\
		(6$\rightarrow$1) & 6-1 &  1 \\ 
		(7$\rightarrow$1) & 7-1 &  1 \\
		(8$\rightarrow$1) & 8-1 &  1 \\
		(9$\rightarrow$1) & 9-8-1 &  3 \\
		(10$\rightarrow$1) & 10-2-1 &  3 \\
		(11$\rightarrow$1) & 11-6-1 &  3 \\		
		\hline
	\end{tabular}  
\end{table}

\begin{table}[!ht]
	\caption{Routing and Wavelength Assignment Information in Optical-computing-enabled Architecture}
	\label{tab:result3}
	\centering
	\begin{tabular}{ccc}
		\hline
		Traffic Demands & Routing & $\lambda$ \\
		\hline
		(2$\rightarrow$1) & 2-1 &   1 \\
		(3$\rightarrow$1) & 3-2-1 & 1 \\
		(4$\rightarrow$1) & 4-10-2-1 &  2 \\
		(5$\rightarrow$1) & 5-7-1 &  1 \\
		(6$\rightarrow$1) & 6-1 &  1 \\ 
		(7$\rightarrow$1) & 7-1 &  1 \\
		(8$\rightarrow$1) & 8-1 &  1 \\
		(9$\rightarrow$1) & 9-8-1 &  1 \\
		(10$\rightarrow$1) & 10-2-1 &  2 \\
		(11$\rightarrow$1) & 11-6-1 &  1 \\		
		\hline
	\end{tabular}  
\end{table}

Next, we are interested in the difference between the optical-computing-enabled and optical-bypass framework in accommodating traffic demands. Table \ref{tab:result2} and Table \ref{tab:result3} showcase the routing and wavelength assignment for all traffic demands arriving at node 1 for both two approaches. It is noted that the routing of all traffic demands remain quite the same for both approaches, except for the traffic demand 3$\rightarrow$1 and while the optical-bypass design needs $three$ wavelengths, the optical-computing design simply requires $two$ wavelengths. Such saving is possible thanks to the optimal interaction of lightpaths in optical-computing network by exploiting the optical aggregation operation. Specifically, in order to gain $one$ wavelength saving, $five$ optical aggregations have been performed between transitional lightpaths as shown in Table \ref{tab:result4}. Note that different from the optical-bypass case, in optical-computing-enabled network, due to the computing operations between lightpaths, it gives rise to the existence of the so-called integrated / computed lightpaths (e.g., $(2 \rightarrow 1) + (3 \rightarrow 1)$). Network design algorithms for optical-computing network must therefore find the routing and wavelength assignment for such newly arisen lightpaths in order to tame spectral benefits as highlighted in Table \ref{tab:result4}.

\begin{table*}[ht]
	\caption{Aggregation Assignment Information (A-Node: Aggregation Node, R-Aggregated Lightpath: Route of the Aggregated Lightpath, W-Aggregated Lightpath: Wavelength Assigned for the Aggregated Lightpath)}
	\label{tab:result4}
	\centering
	\begin{tabular}{cccc}
		\hline
		Aggregation of Demands & A-Node & R-Aggregated Lightpath & W-Aggregated Lightpath \\
		\hline
		$(2 \rightarrow 1) + (3 \rightarrow 1)$ & 2 & (2-1) & 1 \\
		$(4 \rightarrow 1) + (10 \rightarrow 1)$ & 10 & (10-2-1) & 2 \\
		$(5 \rightarrow 1) + (7 \rightarrow 1)$ & 7 & (7-1) & 1 \\	
		$(6 \rightarrow 1) + (11 \rightarrow 1)$ & 6 & (6-1) & 1 \\
		$(8 \rightarrow 1) + (9 \rightarrow 1)$ & 8 & (8-1) & 1 \\
		\hline
	\end{tabular}  
\end{table*}

\section{Conclusions}
This paper has identified for the first time the critical bottleneck of optical-bypass networking, that is, transitional lightpaths crossing the same intermediate nodes are separated from each other, usually in either time, frequency or spatial domain in avoidance of adversarial interference. In reversing this long-established foundation, we have proposed a new architectural paradigm, entitled, optical-computing-enabled network, by exploiting the possibility for superposition of in-transit lightpaths to achieve greater capacity efficiency. In demonstrating the efficacy of our proposal, a detailed case study focusing on using optical aggregation operation merging two QPSK channels into a single 16-QAM one has been presented and formulated in the form of an integer linear programming model. Our formulation has optimally determined pair of demands for aggregation, the node at which the aggregation takes place and the wavelength selection for aggregated lightpaths. Numerical results on COST239 topology has been provided to quantify the efficacy of the optical-computing-enabled approach versus the conventional optical-bypass-enabled one. \\

Albeit still in early stages, the prospect of superimposing transitional optical channels heralds a transformation for optical switching architectures from conventional functions of add-drop and cross-connect to pro-actively performing various computing operations on optical channels in the photonic domain. This is set to have massive impacts for optical networks toward greater spectral, cost and energy efficiency, driving more synergies between photonic technologies, systems and computing / networking architectures. From a network design and planning perspective, the potential benefits of optical-computing networking paradigm, however, come at the expenses of, among others, more complicated network design problems, which are generally at least one order of magnitude computationally harder than counterparts in optical-bypass schemes. This therefore necessitates a radically different framework in network design algorithms which may collectively be referred as \textit{optical network design and planning 2.0} to differentiate itself from the well-developed algorithms that have long been successfully used in optical-bypass framework. Furthermore, the arrival of integrated / computed lightpaths in optical-computing-enabled network brings about new challenges to be addressed for control plane protocols, particularly concerning tracking and distinguishing them from ordinary ones as they are routed through the network. Backward-compatibility is also a critical issue to be addressed in future works and among others, the possibility of re-purposing already installed transponders for computing purposes  remains to be elucidated. \\

\bibliographystyle{elsarticle-num}

\bibliography{oft_final}

\end{document}